\documentclass[prb,showpacs,preprintnumbers,amsmath,amssymb,twocolumn]{revtex4-1}
\usepackage{graphicx}
\usepackage{dcolumn}
\usepackage{bm}
\usepackage{epstopdf}
\usepackage{natbib}
\begin{document}
`
\title{Temperature - Pressure phase diagram of the cubic Laves phase Au$_2$Pb}

\author{K. W. Chen$^{1,2}$, D. Graf$^1$, T. Besara$^1$, A. Gallagher$^{1,2}$, N. Kikugawa$^{1,4}$, L. Balicas$^1$, T. Siegrist$^{1,3}$, A. Shekhter,$^1$ R. E. Baumbach$^1$
}
\affiliation{$^1$National High Magnetic Field Laboratory, Florida State University}
\affiliation{$^2$Department of Physics, Florida State University}
\affiliation{$^3$Department of Chemical and Biomedical Engineering, Florida State University}
\affiliation{$^4$National Institute for Materials Science, 3-13 Sakura, Tsukuba, Japan}
\date{\today}

\begin{abstract}
The temperature ($T$) as a function of pressure ($P$) phase diagram is reported for the cubic Laves phase compound Au$_2$Pb, which was recently proposed to support linearly dispersing ``topological" bands, together with conventional quadratic bands. At ambient pressure, Au$_2$Pb exhibits several structural phase transitions at $T_1$ $=$ 97 K, $T_2$ $=$ 51 K, and $T_3$ $=$ 40 K with superconductivity below $T_{\rm{c}}$ $=$ 1.2 K. Applied pressure results in a rich phase diagram where $T_1$, $T_2$, and $T_3$ evolve strongly with $P$ and a new phase is stabilized for $P$ $>$ 0.64 GPa that also supports superconductivity below 1.1 K. These observations suggest that Au$_2$Pb is an ideal system in which to investigate the relationship between structural degrees of freedom, band topology, and resulting anomalous behaviors.
\end{abstract}

\maketitle

\section{Introduction}
The cubic Laves phase compounds $M_2$$X$ can be considered as being relatives of the pyrochlore structure, where the tetrahedrally arranged $M$ atoms are distributed inside a face centered cubic arrangement of $X$ atoms.~\cite{Pearson} Prior work has shown that this structure accommodates a multitude of binary chemical combinations,~\cite{stein2004} making it a deep reservoir for electronic and magnetic states including superconductivity,\cite{Finlayson1978,Huxley1993,Saini2004} electronic valence changes,\cite{Barberis1980,Felner1986} itinerant electron magnetism,\cite{Matthias1958} geometrically frustrated magnetism,\cite{Shiga1994} and other effects. Recent work addressing Au$_2$Pb indicates that this family of materials is now expanded to include topologically protected electronic states. The calculated electronic band structure for Au$_2$Pb includes a crystallographically protected Dirac point at room temperature, in addition to other conventional bands at the Fermi energy ($E_{\rm{F}}$).~\cite{Schoop2015} With decreasing temperature, Au$_2$Pb undergoes several structural phase transitions at $T_1$ $=$ 97 K, $T_2$ $=$ 51 K, and $T_3$ $=$ 40 K. Below $T_3$, Au$_2$Pb enters a primitive orthorhombic structure, where electronic structure calculations predict that the Dirac point is gapped but retains topological protection. Other band crossings at $E_F$ are also present at low $T$, providing conduction electrons that produce metallic behavior and conventional BCS superconductivity ($T_c$ $\approx$ 1.2 K). 

While the presence of conventional bands at $E_F$ makes Au$_2$Pb distinct from simple three dimensional Dirac semimetals (at high $T$)~\cite{Young2012} or topological insulators (at low $T$)~\cite{Zhang2009}, it is of interest because it offers the opportunity to study the relationship between conventional electrons and quasiparticles associated with linearly dispersing bands. This includes the question of how conventional BCS superconductivity might interact with gapped topological bands. It is also possible that the simultaneous presence of linearly dispersing and quadratic bands could result in unusual transport and thermodynamic properties. The complex structural evolution additionally allows for study of how a Dirac point is influenced by changes in the crystal symmetry. 

Hydrostatic pressure is a useful tool to systematically probe lattice structures and concomitant electronic states. For this reason, we undertook to study electrical transport under applied pressure $P$ $<$ 1.8 GPa in single crystals of Au$_2$Pb. Our measurements reveal an unusually rich $T-P$ phase diagram that includes several structural phases and superconducting regions. In particular, we find that while $T_1$increases monotonically for pressures up to 1.83 GPa, $T_2$ and $T_3$ are initially suppressed and meet near $P_1$ $\approx$ 0.64 GPa. For $P$ $\geq$ $P_1$ $\approx$ 0.64 GPa, another phase boundary that likely is structural in nature is seen at $T_4$, which increases with $P$. Superconductivity is seen at low temperatures throughout the entire measured $P$ range. This multitude of structural phases and the persistence of superconductivity will be useful for uncovering anomalous electronic behavior and disentangling topological behaviors from those that are due to material specific details. We additionally compare to the Laves phase analogue, Au$_2$Bi, which is a conventional metal that does not exhibit any structural phase transitions but does support superconductivity.

\section{Experimental Methods}
Single crystals of Au$_2$Pb were grown using a molten Pb flux as described previously \cite{Schoop2015}. 
Elemental Au and Pb were mixed in the ratio 2:3 and sealed under vacuum in a quartz tube.
The mixture was heated at a rate of 50 $^{\rm{o}}$C/hr to 600 $^{\rm{o}}$C, held at this temperature for 24hr, and slowly cooled at a rate of 3$^{\rm{o}}$C/hr to 300$^{\rm{o}}$C. 
The crystals were separated from the flux by centrifuging at this temperature. Single crystals of Au$_2$Bi were made using a similar approach, where the molar ratio was 3:7.

Single crystals of Au$_{2}$Pb and Au$_{2}$Bi were structurally characterized at room temperature by single crystal x-ray diffraction using an Oxford-Diffraction Xcalibur2 CCD system with graphite monochromated Mo $K\alpha$ radiation. 
Data was collected using $\omega$ scans with 1$^{\rm{o}}$ frame widths to a resolution of 0.5 \textrm{\AA}, equivalent to $2\theta$ = 90$^{\rm{o}}$. 
The data collection, indexation, and absorption correction were performed using the Agilent CrysAlisPro software \cite{CrysAlisPro}. Subsequent structure refinement was performed by using CRYSTALS \cite{Crystals}. 
 A crystallographic information file (CIF) for Au$_2$Pb has been deposited with ICSD (CSD No. 430187)~\cite{ICSD}. 
 For subsequent measurements, single crystals were selected and aligned using a four-axis Enraf Nonius CAD-4 Single Crystal X-Ray Diffractometer. 
 The obtained orientation matrix allowed for an unambiguous determination of the crystalline axes to within a degree.

Specimens were  prepared for electrical resistivity $\rho$ measurements by attaching Pt wires using a home built micro-spot welder. Samples of Au$_2$Pb were loaded into a piston-cylinder pressure cell. 
The pressure $P$ was determined by measuring the superconducting transition of 6N polycrystalline Pb. Daphne 7474 oil was used as the pressure-transmitting medium~\cite{Murata1997,Murata2008}. The electrical resistivity $\rho$ was measured with currents applied along the [100] direction (at room temperature) for temperatures 1.8 K $<$ $T$ $<$ 300 K and $P$ $\leq$ 1.8 GPa using a Quantum Design Physical Property Measurement System and an SR830 lock-in amplifier with a Keithley 6221 as the current source. 
Additional resistivity measurements under applied pressure were performed at the National High Magnetic Field Laboratory, Tallahassee, for temperatures between 500 mK $<$ $T$ $<$ 2 K using a standard He-3 cryostat. Electrical resistivity and heat capacity measurements were performed at ambient pressure on an aligned single crystal of Au$_2$Bi using the He-3 option in a Quantum Design Physical Properties Measurement System for temperatures 400 mK $<$ $T$ $<$ 20 K.

\section{Results}
\begin{figure}[!t]
    \begin{center}
        \includegraphics[width=3.0in]{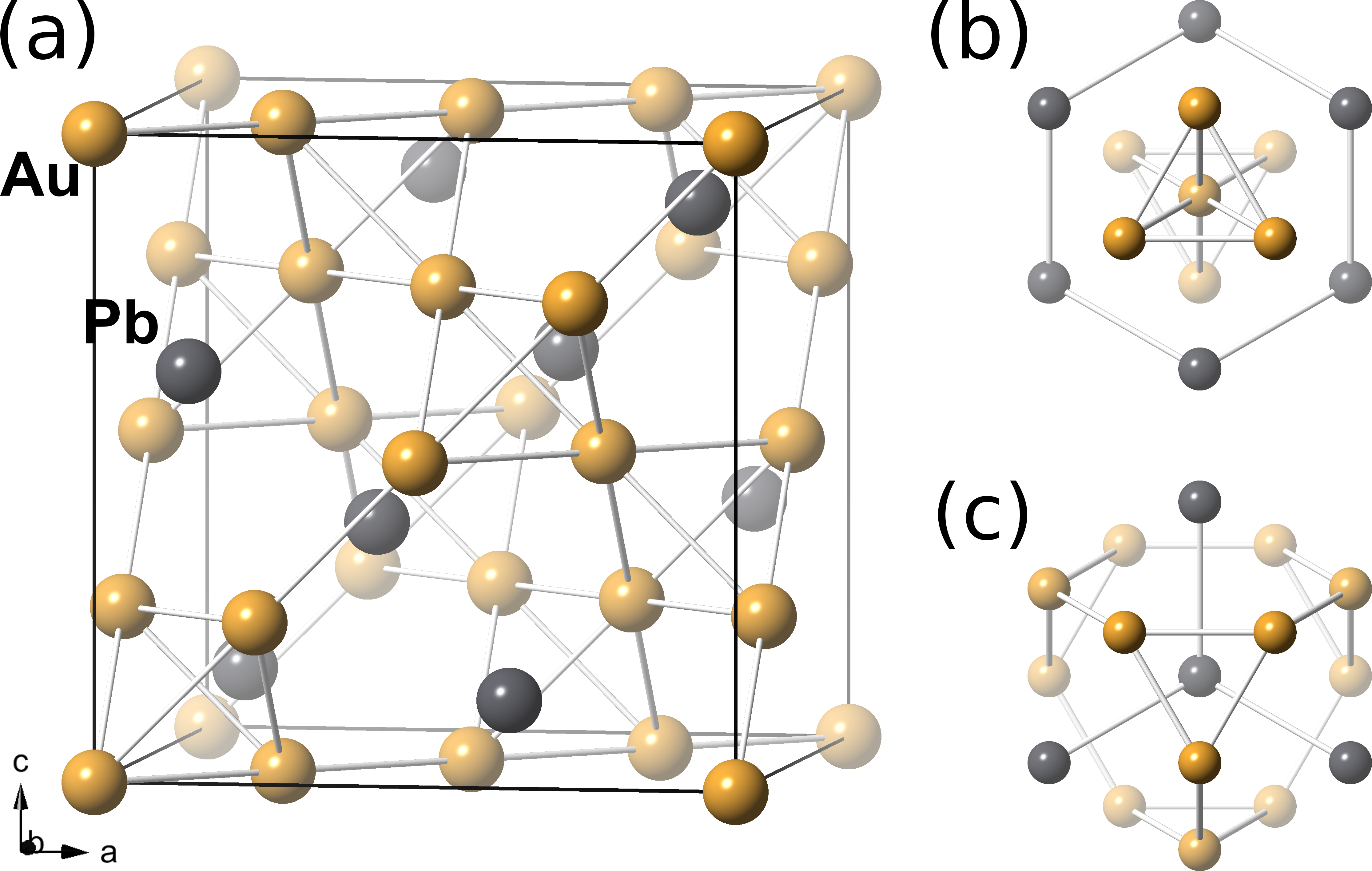}
        \caption{Crystal structure of Au$_{2}$Pb, showing (a) the unit cell, (b) local coordination of Au along [111], and (c) local coordination of Pb along [111].}
        \label{fig:struct}
    \end{center}
\end{figure}

Our measurements verify that Au$_{2}$Pb and Au$_{2}$Bi crystallize in the space group $Fd\bar{3}m$ (\#227) where Au$_2$Pb has unit cell parameter $a=7.9141(2)$~\textrm{\AA} (Fig. \ref{fig:struct}(a)); the structural parameters resulting from the x-ray refinement for Au$_2$Pb are summarized in Table \ref{tbl:xray}.
 Au$_{2}$Pb and Au$_{2}$Bi are cubic Laves phases with a pyrochlore lattice,~\cite{Pearson} where the gold is in the 16c position at $(0,0,0)$ and lead/bismuth are in the 8b position at $(3/8,3/8,3/8)$ (Table \ref{tbl:coord}).
  The single crystal x-ray diffraction for Au$_2$Pb displayed a site vacancy defect on the Pb-site, yielding a formal stoichiometry of Au$_{2}$Pb$_{0.95}$. 
  For simplicity, the compound will be referred to as Au$_{2}$Pb throughout the rest of the manuscript.

The gold atoms form a network of corner-shared tetrahedra connected along the $\langle111\rangle$-directions, with a Au\textendash Au distance of 2.798~\textrm{\AA}.
The gold network in turn forms voids in which the lead atoms reside as trigonal chains along $\langle110\rangle$, with a Pb\textendash Pb distance of 3.427~\textrm{\AA}. 
Au is 12-fold coordinated, surrounded by six Au-atoms and six Pb-atoms (Fig. \ref{fig:struct}(b)), forming a icosahedron. 
Pb is 16-fold coordinated with twelve Au-atoms and four Pb-atoms surrounding it (Fig. \ref{fig:struct}(c)), forming a 16-vertex Frank-Kasper polyhedron~\cite{Pearson}.

\begin{table}[!b]
    \begin{center}
        \caption[]{Selected single crystal x-ray diffraction data, along with collection and refinement parameters. (Lattice parameters are for Au$_{2}$Pb$_{0.951(11)}$) }
        \begin{tabular}{l l}
            \hline
            \textbf{Compound} & Au$_{2}$Pb \\
            \hline
            Composition & Au$_{2}$Pb$_{0.951(11)}$ \\
            Formula weight & 590.90 g/mol \\
            Space group & $Fd\bar{3}m$ (\#227)\\
            Unit cell parameter & $a=7.9141(2)$ \textrm{\AA} \\
            Volume & 495.69(1) \textrm{\AA}$^{3}$ \\
            $Z$ & 8 \\
            Data collection range & $4.46^o\leq\theta\leq45.93^o$ \\
            Reflections collected & 6344 \\
            Independent reflections & 130 \\
            Parameters refined & 6 \\
            $R_{1}$, $wR_{2}$ & 0.0431, 0.0687 \\
            Goodness-of-fit on $F^{2}$ & 0.9942 \\
            \hline
        \end{tabular}
        \label{tbl:xray}
    \end{center}
\end{table}

\begin{table}[!b]
    \begin{center}
        \caption[]{Atomic coordinates and equivalent thermal displacement parameters, along with interatomic distances.}
        \begin{tabular}{l l l l l l l}
            \hline
            Atom & Site & Occ. & $x$ & $y$ & $z$ & $U_{\textrm{eq}}$ (\textrm{\AA}$^{2}$) \\
            \hline
            Au & 16c & 1 & 0 & 0 & 0 & 0.0320(5) \\
            Pb & 8b & 0.951(11) & 3/8 & 3/8 & 3/8 & 0.0191(5) \\
            \hline
            \\
            \hline
            \multicolumn{3}{l}{Bond} & \multicolumn{3}{l}{Distance (\textrm{\AA})} \\
            \hline
            \multicolumn{3}{l}{Au\textemdash Au} & \multicolumn{3}{l}{2.798(1)} \\
            \multicolumn{3}{l}{Pb\textemdash Pb} & \multicolumn{3}{l}{3.427(1)} \\
            \multicolumn{3}{l}{Au\textemdash Pb} & \multicolumn{3}{l}{3.281(1)} \\
            \hline
        \end{tabular}
        \label{tbl:coord}
    \end{center}
\end{table}

\begin{figure}[!tht]
    \begin{center}
        \includegraphics[width=3.4in]{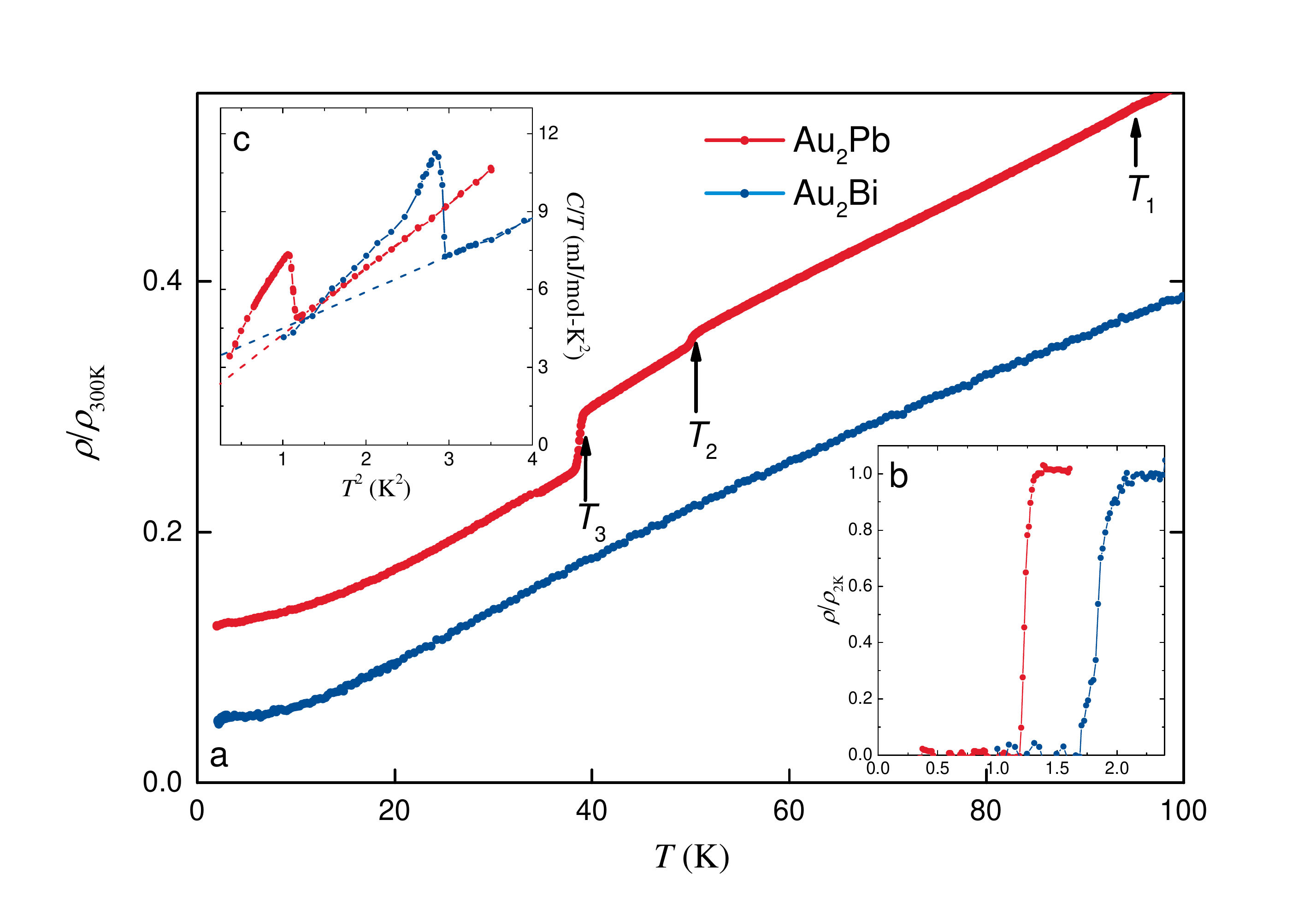}
        \caption{(a) Cooling curves of the room temperature normalized electrical resistivity $\rho$/$\rho_{\rm{300K}}$ versus temperature $T$ for Au$_2$Pb and Au$_2$Bi, where the structural transitions $T_1$, $T_2$, and $T_3$ in Au$_2$Pb are indicated by arrows. 
        (b) Zoom of the normalized resistivity $\rho$/$\rho_{\rm{2K}}$ for $T$ $\leq$ 2.2 K, where superconductivity appears as sharp drops to zero resistivity. 
        (c) Heat capacity divided by temperature $C/T$ vs. $T^2$ for Au$_2$Pb and Au$_2$Bi, where the superconducting transitions appear as sharp lambda-type transitions. The dotted lines are fits to the data as described in the text.}
        \label{fig:rho}
    \end{center}
\end{figure}

The ambient pressure electrical resistivities for Au$_2$Pb and Au$_2$Bi are shown in Fig.~\ref{fig:rho}. The residual resistivity ratios $RRR$ $=$ $\rho_{300K}$/$\rho_{0}$ for Au$_2$Pb and Au$_2$Bi are 9 and 20, respectively. 
For Au$_2$Pb, the structural transitions at $T_2$ $=$ 51 K and $T_3$ $=$ 40 K are easily observed and superconductivity appears below $T_{\rm{c}}$ $=$ 1.2 K (Fig.~\ref{fig:rho}b), as previously reported.~\cite{Schoop2015} 
In comparison, Au$_2$Bi shows typical metallic behavior with no evidence for structural phase transitions, suggesting that it is useful as a conventional metallic Laves phase analogue to Au$_2$Pb. 
We additionally find that Au$_2$Bi exhibits superconductivity near 1.7 K (Fig.~\ref{fig:rho}b), as previously reported.~\cite{Khan1975} Heat capacity measurements (Fig.~\ref{fig:rho}c)  
show that the Au$_2$Bi superconductivity occurs in the bulk, where the ratio $\Delta$$C$/$\gamma$$T_c$ $=$ 1.9, is comparable to what is seen for Au$_2$Pb.~\cite{Schoop2015} 
This value is larger than what is expected for a BCS superconductor ($\Delta$$C$/$\gamma$$T_c$ $=$ 1.43), but is smaller than that of elemental lead.~\cite{Neighbor1967} 
A fit to the data using the expression $C/T$ $=$ $\gamma$ $+$ $\beta$$T^2$ gives an electronic coefficient $\gamma$ $=$ 2.3 mJ/mol-K$^2$ and a Debye temperature $\Theta_{\rm{D}}$ $=$ 153 K ($\beta$ $=$ 1.6 mJ/mol-K$^4$). 
For Au$_2$Pb, we find $\gamma$ $=$ 1.9 mJ/mol-K$^2$ and $\Theta_{\rm{D}}$ $=$ 133 K, as previously reported.~\cite{Schoop2015}

\begin{figure}[!tht]
    \begin{center}
        \includegraphics[width=3.4in]{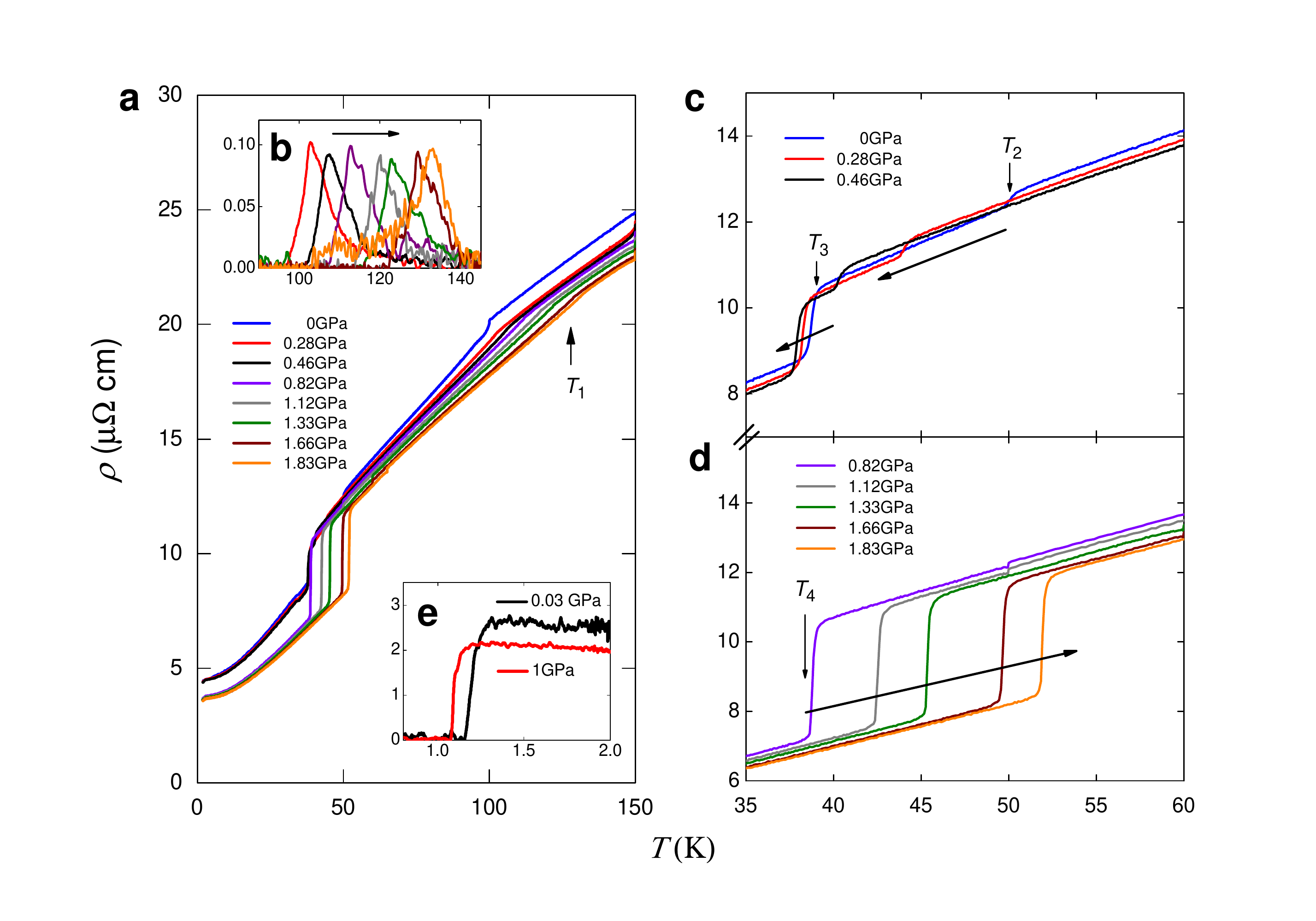}
        \caption{(a) Cooling curves of the electrical resistivity $\rho$ versus temperature $T$ at pressures 0 GPa $\leq$ $P$ $\leq$ 1.83 GPa for Au$_2$Pb. Note that the sharp kink for $P$ $=$ 0 near $T$ $=$ 100 K is an artefact due to a change in the $T$ sweep rate in this data set only.  
        (b) Zoom of $\rho(T)$ at several $P$ near the anomaly $T_1$, after background subtraction. 
        The arrow is a guide to the eye showing the transition temperatures and their evolution. (c) Zoom of $\rho(T)$ for $P$ $\leq$ 0.46 GPa at temperatures near the sharp anomalies at $T_2$ and $T_3$. 
        (d) Zoom of $\rho(T)$ for $P$ $\geq$ 0.82 GPa at temperatures near the sharp anomaly at $T_4$. (e) $\rho(T)$ for $P$ $=$ 0.03 and 1 GPa showing superconductivity at low $T$. 
        The superconducting temperature evolves from $T$$_c$ $=$ 1.2 K at 0.03 GPa to $T$$_c$ $=$ 1.1 K at 1 GPa.}
        \label{fig:rho_T}
    \end{center}
\end{figure}

The electrical resistivity as a function of temperature under several values of hydrostatic pressure is shown in Fig.~\ref{fig:rho_T}. 
Metallic behavior is seen for all pressures with a room temperature resistivity of 35.9 $\mu$$\Omega$cm for $P$ $=$ 0 which decreases to 33.0 $\mu$$\Omega$cm for $P$ $=$ 1.83 GPa. We find residual resistivity ratios $RRR$ $=$ $\rho_{295K}$/$\rho_0$ in the range 8.1-9.2, slightly increasing with increasing $P$. 
The previously reported transition at $T_1$ appears as a weak reduction in $\rho$, 
which becomes more pronounced and increases with increasing $P$ at a rate of 19 K/GPa (Fig.~\ref{fig:rho_T}b), 
indicating that this phase transition would occur at room temperature near $P$ $\approx$ 9 GPa. 
$T_2$ and $T_3$ appear as sharp reductions in the resistivity, consistent with an earlier report of structural phase transitions at these temperatures under ambient pressure (Fig.~\ref{fig:rho_T}a,d,e).\cite{Schoop2015}  
The features at $T_2$ and $T_3$ decrease linearly at rates of 21.5 K/GPa and 1.8 K/GPa, respectively, and finally intersect near $P_1$ $\approx$ 0.64 GPa (determined by extrapolation). 
For pressures larger than 0.64 GPa, another phase transition appears as a sharp reduction in the resistivity at $T_4$, which produces a more pronounced decrease in resistivity than $T_2$ and $T_3$ combined. 
Unlike $T_2$ and $T_3$, $T_4$ increases linearly at a rate of 12.6 K/GPa, reaching 52 K at $P$ $=$ 1.83 GPa (Fig.~\ref{fig:rho_T}d). All of the observed phase transitions show hysteresis under $P$ (summarized in Fig~\ref{fig:phase-diagram}). At low temperature and zero pressure, bulk superconductivity is observed below $T_{\rm{c}}$ $=$ 1.2 K. Superconductivity persists at 1 GPa, with a slightly reduced transition temperature $T_{c}$ $=$ 1.1 K (Fig.~\ref{fig:rho_T}e). 

\begin{figure}[!tht]
    \begin{center}
        \includegraphics[width=3in]{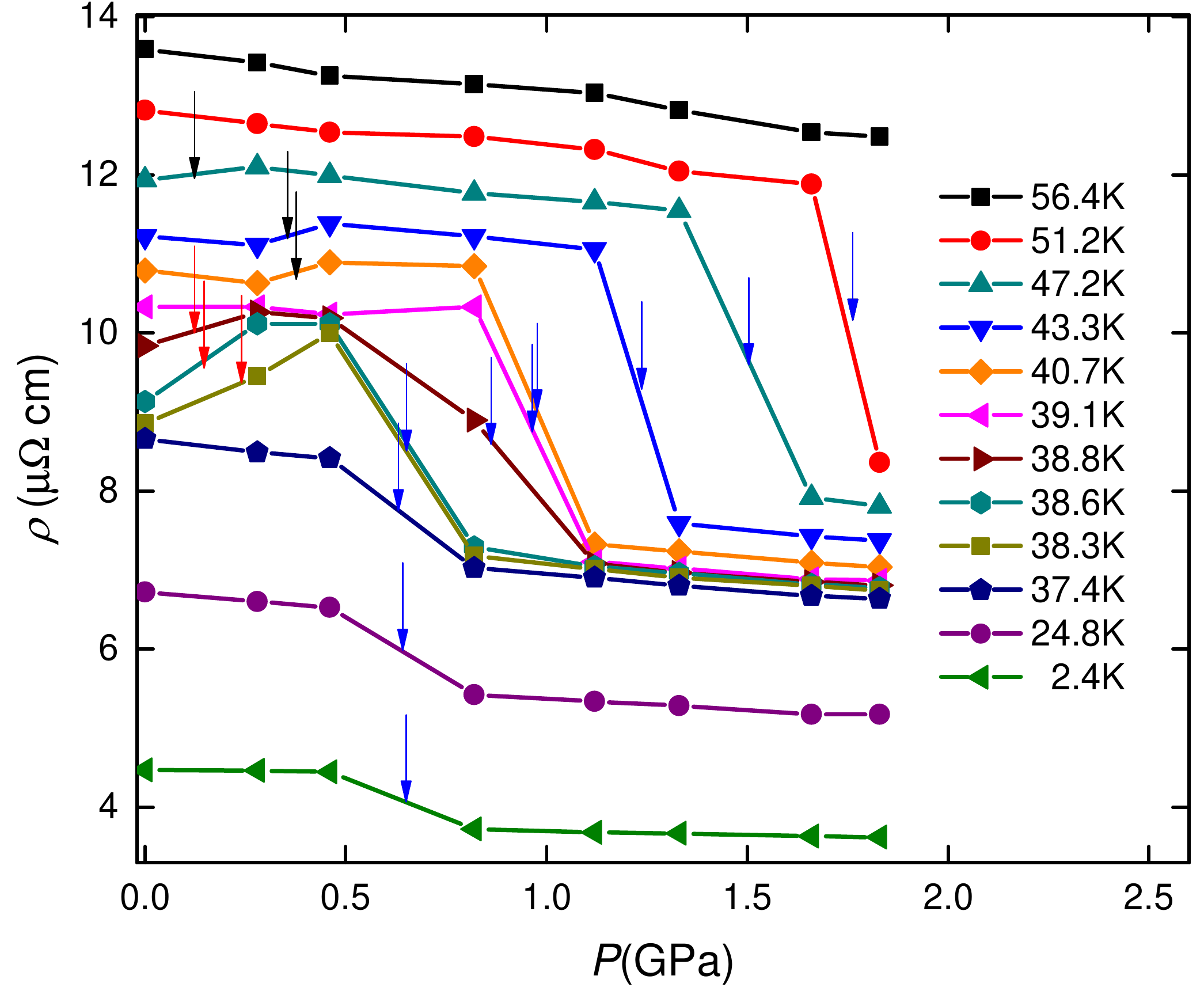}
        \caption{Electrical resistivity $\rho$ versus pressure $P$ at several temperatures $T$. The arrows indicate the phase transitions.  Several structural transitions appear as sharp kinks in $\rho(P)$ as described in the text. The black arrows indicate $T_2$, the red arrows indicate $T_3$, and the blue arrows indicate $T_4$.}
        \label{fig:rho_P}
    \end{center}
\end{figure}

\begin{figure}[!tht]
    \begin{center}
        \includegraphics[width=3.25in]{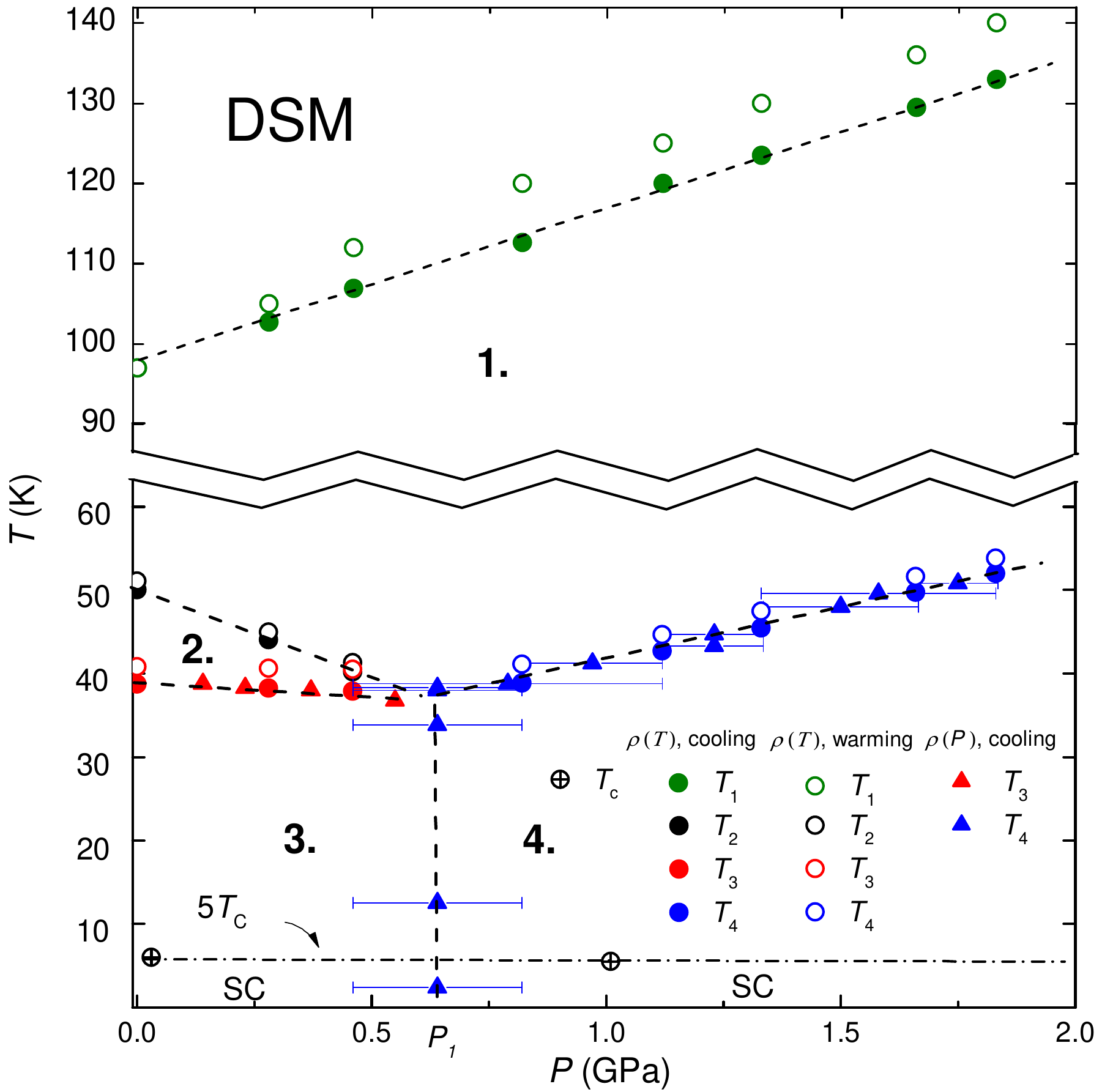}
        \caption{The temperature-pressure $T-P$ phase diagram for Au$_2$Pb. The filled/hollowed circles are from the cooling/warming curves of $\rho(T)$ (Fig.~\ref{fig:rho_T}) and the triangles are from the $\rho(P)$ at various $T$ (Fig.~\ref{fig:rho_P}). The ambient pressure phase transitions $T_1$(=97K) for both cooling and warming curves are from ref.~[\onlinecite{Schoop2015}].}
        \label{fig:phase-diagram}
    \end{center}
\end{figure}

Plots of the resistivity at constant temperature under varying pressure $\rho(P)$ are shown in Fig.~\ref{fig:rho_P}. For $T$ $>$ 51 K, $\rho(P)$ is nearly constant for $P$ $<$ 1.8 GPa. For $T$ $\leq$ 51 K, there is a large reduction of $\rho$ which first appears at large $P$ and $T$ and moves to lower $P$ with decreasing $T$. For 38 K $<$ $T$ $<$ 51 K, there is a second transition at lower pressures that evolves with temperature. For $T$ $<$ 38 K, there is a single transition pressure that remains nearly constant at $P_1$ $\approx$ 0.64 GPa with decreasing $T$. We infer that these features are associated with the low temperature structural phase transitions seen in $\rho(T)$.

These results are collected in Fig.~\ref{fig:phase-diagram} to build the $T-P$ phase diagram. The solid circles are the $\rho(T)$ cool down curves (Fig.~\ref{fig:rho_T}), the hollow circles are from the $\rho(T)$ of warm up curves and the triangles are from $\rho(P)$ (Fig.~\ref{fig:rho_P}). While the precise location of the phase boundaries in the phase diagram need additional measurements to
establish the extent of the phase fields (particularly where $T_2$ and $T_3$ intersect), there appear to be five distinct regions including the previously described Dirac semimetal phase which extends from room temperature to $T$ $=$ 97 K for $P$ $=$ 0 and several other low $T$ phases labeled 1-4. Region 1 spans the entire pressure range. Region 2 is a small wedge separating 1 and 3. Regions 3 and 4 are the low $T$/low $P$ and low $T$/high $P$ phases which are separated by the nearly vertical phase boundary at $P_1$ $\approx$ 0.64 GPa. We further find that superconductivity is present in both regions 3 and 4, where the transition temperature is slightly suppressed by $\approx$ 0.1 K in region 4.

\section{Discussion}
The past several years have witnessed intense interest in materials that support novel electronic states arising from unusual band structures. Examples include materials with topologically protected linearly dispersing bands such as Dirac semimetals (Na$_3$Bi~\cite{Liu2014} and Cd$_3$As$_2$~\cite{Neupane2014}), Weyl semimentals (Nb,Ta)(P,As),~\cite{Huang2015}  and topological insulators (Bi$_2$(Se,Te)$_3$),~\cite{Zhang2009} as well as other unusual metallic phenomena such as the axial anomaly (PdCoO$_2$~\cite{kikugawa}). While these materials exhibit remarkable behavior, they are often limited in terms of opportunities to test microscopic models by transforming their thermodynamic ground state: e.g., using probes such as applied pressure. As demonstrated in this study, Au$_2$Pb and potentially the alloy Au$_2$Pb$_{1-x}$Bi$_x$ provide the unusual examples where the evolution of topological behavior under the influence of intrinsic structural phase transitions can be systematically explored. In order to do this, it will first be important to determine the structures and electronic characteristics of regions 1, 2, and 4 and to carry out accompanying band structure calculations.

The presence of superconductivity in both the low and high pressure regions for Au$_2$Pb presents additional opportunities. Prior work suggests that the ambient pressure superconductivity in Au$_2$Pb occurs in the bulk and can be described as being due to weak coupling electron-phonon interactions in the BCS framework. It is appealing to attribute this superconductivity to the quadratically dispersing bands that are predicted to cross $E_{\rm{F}}$ along the $\Gamma$ $-$ $K$ line: i.e., the ``conventional bands".~\cite{Schoop2015}  If this is the right picture, then Au$_2$Pb is a model system in which to investigate what happens in a material that hosts both conventional superconductivity and gapped linearly dispersing bands. Prior work suggests that such a material might support Majorana fermions,~\cite{Leijnse2012} although this has yet to be substantiated. Comparison to Au$_2$Bi will help to independently determine whether behavior seen in Au$_2$Pb is uniquely associated with topological bands. 

Our pressure study further amplifies the novelty of this situation by showing that there is a distinct high pressure phase which also supports superconductivity. It will be of interest to determine whether the high pressure phase includes topological bands and whether they are significantly different from those at ambient pressure. In the case that the high pressure bands are all conventional, this would give the opportunity to definitively separate what superconducting behaviors are uniquely associated with the presence of topological bands. In particular, further studies of the magnetoresistance and magnetization will be of use to search for unusual power law behaviors and to index quantum oscillation Landau levels, which would unambiguously characterize topological behavior.

\section{Conclusion}

We have presented the $T-P$ phase diagram for the cubic Laves phase compound Au$_2$Pb, which supports several structural phases and superconductivity over a broad range of pressures. At ambient pressure, we compare it to the conventional metallic Laves phase analogue Au$_2$Bi. Our observations have important implications for understanding the interplay between conventional and linearly dispersing bands, particularly with respect to possible anomalous transport behavior and the relationship with superconductivity. Owing to its structural tunability, both with temperature and pressure, it will be possible to systematically explore these effects under different conditions. Such information is needed to separate what features should be attributed to topological behavior and which are merely due to specific details of this material. 

\section{Acknowledgements}
This work was performed at the National High Magnetic Field Laboratory (NHMFL), which is supported by National Science Foundation Cooperative Agreement No. DMR-1157490, the State of Florida and the DOE. T.B. and T.S. are supported by the U.S. Department of Energy, Office of Basic Energy Sciences, Materials Sciences and Engineering Division, under Award DE-SC0008832. L.B. is supported by DOE-BES through award DESC0002613. N.K. acknowledges the support of overseas researcher dispatch program in NIMS.

\end{document}